\documentstyle[12pt]{article}
\input epsf

\begin{document}
\begin{titlepage}
\thispagestyle{empty}

%\large

%\begin{center}

\title{ 
\vspace*{-1cm}
\begin{flushright}
{\small CPHT S758.0100}
\end{flushright}
\vspace{2.0cm}
Dispersion Relation Analyses of Pion Form Factor,  Chiral Perturbation Theory
and Unitarized Calculations }
\vspace{4.0cm}
\author{Tran N. Truong \\
\small \em Centre de Physique Th{\'e}orique, 
{\footnote {unit{\'e} propre 014 du
CNRS}}\\ 
\small \em Ecole Polytechnique \\
\small \em F91128 Palaiseau, France}

\date{February 2000 }

\maketitle
%\end{center}

\begin{abstract}
The Vector Pion form factor below 1 GeV is analyzed using  experimental data on its 
modulus, the P-wave pion pion phase shifts and dispersion relation. It is found that
causality is satisfied. Using dispersion relation, terms proportional to $s^2$ and $s^3$ are
calculated using the experimental data, where $s$ is the momentum transfer. They are much
larger than the one-loop and two-loop Chiral Perturbation   Theory calculations.  Unitarized
model  calculations agree very well with dispersion relation results.

\end{abstract}
\end{titlepage}

%\newpage

 Chiral Perturbation Theory (ChPT) \cite{holstein,Weinberg, GL1, GL2} is a
well-defined perturbative procedure
 allowing one to calculate systematically low energy phenomenon involving
soft pions. It is now widely used to analyze  
low energy pion physics even in the presence
of  resonance as long as the energy region of interest is sufficiently far from the
resonance. In this scheme, the
unitarity relation is satisfied perturbatively order by order.

 The standard
procedure of testing  ChPT calculation of the pion form factor \cite{Gasser3}, which claims
to support the perturbative scheme, is shown here to be unsatisfactory. This is so because 
the calculable terms are extremely small, less than 1.5\% of the uncalculable terms at an
energy of 0.5 GeV or lower whereas the experimental errors are of the
order 10-15\%. The main purpose of this
note is to show how this situation can be dealt with without asking
for a new measurement of the pion form factor with a precision much
better than 1.5\%.

Although dispersion relation (or causality) has been tested to a great accuracy in the
forward pion nucleon and nucleon nucleon  or anti-nucleon scatterings at low and high
energy, there is no such a test for the form factors. This problem is easy to understand. In
the former case, using  unitarity of the S-matrix, one rigourously obtained the optical
theorem relating the imaginary part of the forward elastic amplitude
 to the total cross section which is a measurable quantity. This result together with 
dispersion relation  establish a general relation between the real  and
imaginary parts of the forward amplitude. 

There is no such a rigourous relation, valid to all energy, for the form factor. In low
energy region, the unitarity of the S-matrix in the elastic region gives a
relation between the phase of the form factor and the P-wave pion pion phase shift, namely
they are the same \cite{watson}. Strictky speaking, this region is extended from the two pion
threshold to
$16m_\pi^2$ where the inelastic effect is rigourously absent. In practice, the
region of the validity of the phase theorem can be exended to 1.1-1.3 GeV because  the
inelastic effect is negligible.  Hence, using the  measurements of the
modulus of the form factor and the P-wave phase shifts,  both the real and imaginary parts of
the form factors are known. Beyond this energy, the imaginary part is
not  known. Fortunately for the present purpose of testing of locality (dispersion relation)
and of the validity of the perturbation theory at low energy, thanks to the use of subtracted
dispersion relations, the knowledge of the imaginary part of the form factor beyond 1.3 GeV
is unimportant.

Because the vector pion form factor $V(s)$ is an analytic function with a cut from
$4m_\pi^2$ to
$\infty$, the $n^{th}$ times subtracted dispersion relation for $V(s)$ reads: 
\begin{equation}
V(s)=a_0+a_1s+...a_{n-1}s^{n-1}+ \frac{s^{n}}{\pi}\int_{4m_\pi^2}^\infty
\frac{ImV(z)dz}{z^{n}(z-s-i\epsilon)}
\label{eq:ff1} 
\end{equation}
where $n\geq 0$ and, for our purpose, the series around the origin is
considered. Because of the real analytic property of $V(s)$, it is real below $4m_\pi^2$. By
taking the real part of this equation, 
$ReV(s)$ is related to the principal part of the dispersion integral involving the $ImV(s)$
apart from the subtraction constants $a_n$.

 The
polynomial on the R.H.S. of Eq. (\ref{eq:ff1}) will be referred in the following as the
subtraction constants and the last term on the R.H.S. as the dispersion integral (DI). The
evaluation of DI as a funtion of $s$ will be done later. 
  Notice that
$a_n=V^n(0)/n!$ is the coefficient of the Taylor series expansion for
$V(s)$, where
$V^n(0)$ is the nth derivative of
$V(s)$ evaluated at the origin. The condition for  Eq. (\ref{eq:ff1}) to be valid 
was  that, on the real positive s axis, the
limit $s^{-n}V(s)\rightarrow 0$ as $s\rightarrow \infty$. By the Phragmen Lindeloff
theorem, this limit would also be true in any direction in the complex s-plane  and hence
it is straightforward to prove Eq. (\ref{eq:ff1}). The coefficient $a_{n+m}$ of the Taylor's 
series  is given by:
\begin{equation}
a_{n+m} = \frac{1}{\pi}\int_{4m_\pi^2}^\infty
\frac{ImV(z)dz}{z^{(n+m+1)}}\label{eq:an}
\end{equation}
where $m\geq 0$. The meaning of this equation is clear: under the above stated assumption,
not only the coefficient $a_n$ can be calculated but all other coefficients $a_{n+m}$
can also be calculated. The larger the value of $m$, the more sensitive is the value of
$a_{n+m}$ to the low energy values of $ImV(s)$. In theoretical work such as in ChPT
approach, to be discussed later, the number of subtraction is such that to make the DI
converges.

The elastic unitarity relation for the pion form factor is $
ImV(s)= V(s)e^{-i\delta(s)}sin\delta(s) $ where $\delta(s)$ is the
elastic P-wave pion pion phase shifts. Below the inelastic threshold of $16m_\pi^2$ where
$m_\pi$ is the pion mass,
 $V(s)$ must have the phase of $\delta(s)$ \cite{watson}. It is an
experimental fact that  below $1.3 GeV$ the inelastic effect is very small, hence, to a
good approximation, the phase  of
$V(s)$ is  $\delta$ below this energy scale.

\begin{equation}
ImV(z)  
                        =\mid V(z)\mid\sin\delta(z) \label{eq:ieu} 
\end{equation}
and 
\begin{equation}
ReV(z)  
                        =\mid V(z) \mid\cos\delta(z) \label{eq:reu} 
\end{equation}

where $\delta$ is the strong elastic P-wave $\pi\pi$ phase shifts. Because the real and
imaginary parts are related by  dispersion relation, it is important to know accurately
$ImV(z)$ over a large  energy region. Below 1.3 GeV,
$ImV(z)$ can be determined accurately because the modulus of the vector form factor
\cite{barkov,aleph} and the corresponding P-wave $\pi\pi$ phase shifts are well measured
\cite{proto, hyams, martin} except at very low energy. 

It is possible to estimate the high energy contribution of the dispersion integral by
fitting the asymptotic behavior of the form factor by the expression,
$V(s)=-(0.25/s)ln(-s/s_\rho)$ where $s_\rho$ is the $\rho$ mass squared.

Using Eq. (\ref{eq:ieu}) and Eq. (\ref{eq:reu}), $ImV(z)$ and $ReV(s)$ are determined
directly from experimental data and are shown, respectively, in Fig.1 and Fig.2.

In the following, for definiteness, one assumes $s^{-1}V(s)\rightarrow 0$ as $s\rightarrow
\infty$ on the cut, i.e. $V(s)$ does not grow as fast as a linear function of $s$. This
assumption is a very mild one because theoretical models assume that the form factor
vanishes at infinite energy as $s^{-1}$. In this case, one can write a once subtracted
dispersion relation for $V(s)$, i.e. one sets $a_0=1$ and $n=1$ in Eq. (\ref{eq:ff1}).

From this assumption on the asymptotic behavior of the form factor, the derivatives of the
form factor at $s=0$ are given by Eq. (\ref{eq:an}) with n=1 and m=0. In particular one has:
 
\begin{equation}
 <r_V^2> = \frac{6}{\pi}\int_{4m_\pi^2}^\infty
\frac{ImV(z)dz}{z^2}\label{eq:rms}
\end{equation}
where  the standard definition $V(s) = 1 + \frac{1}{6} <r_V^2>s+c s^2 + d s^3+...$ is used.
Eq.(\ref{eq:rms}) is a sum rule relating the pion rms radius to the magnitude of the time
like pion form factor and the P-wave $\pi\pi$ phase shift measurements.  Using these data,
the derivatives of the form factor are evaluated at the origin:
\begin{equation}
<r_V^2> = 0.45\pm 0.015 fm^2; c = 3.90\pm 0.20 GeV^{-4}; d = 9.70\pm 0.70 GeV^{-6}
\label{eq:rvn}
\end{equation}
where the upper limit of the integration is taken to be $1.7  GeV^2$. By fitting $ImV(s)$
 by the above mentioned asymptotic expression, the
contribution beyond this upper limit is completely negligible. From the 2
$\pi$ threshold to 
$0.56 GeV$ the experimental data on the  the phase shifts are either poor or
unavailable, an extrapolation procedure based on some model calculations to be discussed
later, has to be used. Because of the threshold behavior of the P-wave phase shift, $ImV(s)$
obtained by this extrapolation procedure is small. They contribute, respectively, 5\%,
15\% and 30\% to the $a_1, a_2$ and $a_3$ sum rules. The results of Eq. (\ref{eq:rvn})
change little if the $\pi\pi$ phase shifts below 
$0.56 GeV$ was extrapolated using an effective range expansion and the modulus of the form
factor using a pole or Breit-Wigner formula.

The only experimental data on the derivatives of the form factor at zero momentum
transfer  is the rms radius of the pion,
$r_V^2= 0.439\pm.008 fm^2$ \cite{na7}. This value is very much in agreement with that 
determined from the sum rules. In fact the sum rule for the rms radius gets overwhelmingly
contribution from the $\rho$ resonance as can be seen from Fig.1. The success of the
calculation  of the r.m.s. radius is a first indication that causality is respected and also
that the extrapolation procedures to low energy for the P-wave $\pi\pi$ phase shifts and for
the modulus of the form factor  are legitimate.

Dispersion relation for the pion form factor is now shown to be 
well verified by the data over a wide energy region. Using $ImV(z)$ as given by Eq.
(\ref{eq:ieu}) together with the once subtracted dispersion relation, one can calculate the
real part of the form factor
$ReV(s)$ in the time-like region and also $V(s)$  in the space like region. Because the
space-like behavior of the form factor is not sensitive to the calculation schemes, it will
not be considered here. The result of this calculation is given in Fig.2. As it can be
seen, dispersion relation results are well satisfied by the data.

The i-loop ChPT result can be put into the following form, similar to Eq. (\ref{eq:ff1}):
\begin{equation}
V^{pert(i)}(s)= 1 +a_1s+a_2s^2+...+a_is^i+D^{pert(i)}(s) \label{eq:peri}
\end{equation}
where $i+1$ subtraction constants are needed to make the last integral on the RHS of this
equation converges and
\begin{equation}
DI^{pert(i)}(s)=  \frac{s^{1+i}}{\pi}\int_{4m_\pi^2}^\infty
\frac{ImV^{pert(i)}(z)dz}{z^{1+i}(z-s-i\epsilon)} \label{eq:Dperti}
\end{equation}
with $ImV^{pert(i)}(z)$ calculated by the $ith$ loop perturbation scheme.

Similarly to these equations, the corresponding experimental vector form factor
$V^{exp(i)}(s)$ and $DI^{exp(i)}(s)$  can be constructed using the same subtraction constants
 as in Eq. (\ref{eq:peri}) but with  the imaginary part replaced by $ImV^{exp(i)}(s)$, 
calculated using Eq. (\ref{eq:ieu}).

The one-loop ChPT calculation requires 2 subtraction constants. The first one is given by
the Ward Identity, the second one is proportional to the r.m.s. radius of the pion. In Fig.
1,  the imaginary part of the one-loop ChPT calculation for the vector pion form factor is
compared with the result of the imaginary part obtained from the experimental data. 
It is seen that they differ very much from each other. One  expects therefore that the
corresponding real parts calculated by dispersion relation should be quite different from
each other.

In Fig.2  the full real part of the one loop amplitude is compared 
 with that obtained from experiment. At very low energy
one cannot distinguish the perturbative result from the experimental one due to the
dominance of the subtraction constants. At an energy around $0.56 GeV$ there is a
definite difference between the perturbative result and the experimental data.  This
difference becomes much clearer in Fig. 3 where only the real part of the perturbative DI,
$ReDI^{pert(1)}(s)$,  is compared with the corresponding experimental quantity,
 $ReDI^{exp(1)}(s)$.
 It is seen that even at 0.5 GeV the discrepancy is clear. Supporters of ChPT would argue
that ChPT would not be expected to work at this energy. One would have to go to a  lower
energy where the data became very inaccurate. 
 
This  argument is false as can be seen by comparing  the ratio
$ReDI^{pert(1)}/ReDI^{exp(1)}$. It is seen in Fig. 4 that \emph{
  everywhere} below 0.6 GeV this ratio
differs from unity by a factor of 6-7 due to the presence of  non perturbative effects.

 Similarly to the one-loop calculation, the  two-loop results are plotted in Fig. (1) - Fig.
(4) \cite{Gasser3}.  Although the two-loop result is better than the one-loop calculation,
because more parameters are introduced, calculating  higher loop effects will not  explain
the data.

It is seen that perturbation theory is inadequate for the vector pion form factor even at
very low momentum transfer. This fact is due to the very large value of the pion r.m.s.
radius or a very low value of the $\rho$ mass $s_\rho$ (see below). In order that the
perturbation theory to be valid the calculated term by ChPT should be much larger than the
 non perturbative effect. At one loop, by requiring the perturbative calculation
dominates over the nonperturbative effects at low energy,  one has  
$s_\rho >>\sqrt{960}\pi f_\pi m_\pi =1.3 GeV^2$ which is far from being satisfied by the
physical value of the  $\rho$ mass.

The unitarized models are now examined. 
It has been shown a long time ago that to take into account of the unitarity relation, it is
better to use the inverse amplitude $1/V(s)$ or the Pade approximant method \cite{Truong1,
Truong3}.

The first model is obtained by introducing a zero in the calculated  form factor in the
ref. \cite{Truong1}  to get an agreement with the experimental r.m.s. radius
. The pion form factor is now multiplied by ${1+\alpha s/s_\rho}$ where 
$s_\rho$ is the $\rho$ mass squared
\cite{Truong4}. 

The experimental data can be  fitted with a $\rho$ mass equal to $0.773 GeV$ and
$\alpha=0.14$. These results are in excellent agreement
with the data \cite{aleph,na7}.

The second model, which  is more complete at the expense of introducing  more parameters,
is  based on the two-loop ChPT calculation with unitarity taken into account. It has the
singularity associated with the two loop graphs. Using the same inverse amplitude method
as was done with the one-loop amplitude, but  generalizing this method to two-loop
calculation, Hannah has recently obtained a remarkable fit to the pion form factor in the
time-like and space-like regions. His result is equivalent to  the (0,2)  Pad{\'e} approximant
method as applied to the two-loop ChPT calculation \cite{hannah1}.
  Both models contain ghosts which can be shown to be unimportant
\cite{ht}.

As can be seen from Figs. 1, 2 and 3 the imaginary and real parts of these two models are
very much in agreement with the data. A small deviation of $ImV(s)$  above $0.9 GeV$ is due
to a small deviation of the phases of
$V(s)$ in these two models from the data of the P-wave $\pi\pi$ phase shifts.

In conclusion, higher loop perturbative calculations
 do not solve the unitarity problem. The perturbative
scheme has to be supplemented by the well-known unitarisation schemes such as the inverse
amplitude, N/D and  Pad{\'e} approximant methods \cite{Truong1, Truong3,  hannah1,  ht,
Truong5, lehmann}.

The author would like to thank Torben Hannah for a detailed explanation of his
calculation of the two-loop vector pion form factor and also for a discussion of the
experimental situation on the pion form factor data. Useful conversations with T. N. Pham are
acknowleledged.

\newpage

{\bf Figure Captions}

Fig.1~: The imaginary part of the vector pion form factor $ImV$, given by Eq.
(\ref{eq:ieu}), as a function of  energy in the $GeV$ unit. The solid curve  is the the
experimental results with experimental errors; the long-dashed curve is the two-loop ChPT
calculation, the medium long-dashed curve is the one-loop ChPT calculation, the short-dashed
curve is from the modified unitarized one-loop ChPT calculation fitted to the $\rho$ mass
and the experimental r.m.s. radius, and the dotted curve is the unitarized two-loop
calculation of Hannah
\cite{hannah1}.

Fig. 2~: The real parts of the pion form factor $ReV$, given by Eq. (\ref{eq:reu}) as a
function of energy. The curves are as in Fig. 1. The real part of the
form factor calculated by the once subtracted dispersion relation
using the experimental imaginary part is also given by the solid line.

Fig. 3~: The real parts of the dispersion integral ReDI as a function of energy. The curves are as in Fig. 1.

Fig. 4~:The ratio of the one-loop ChPT to the corresponding experimental quantity,
$ReDI^{pert(1)}/ReDI^{exp(1)}$, defined by Eq. (\ref{eq:Dperti}), as a function of energy, 
is given by the solid line; the corresponding ratio for the two-loop result is given by the
dashed line. The ratio of the unitarized models to the experimental results is unity (not
shown). The experimental errors are estimated to be less than 10\%.
\newpage
\begin{figure} 
\epsfbox{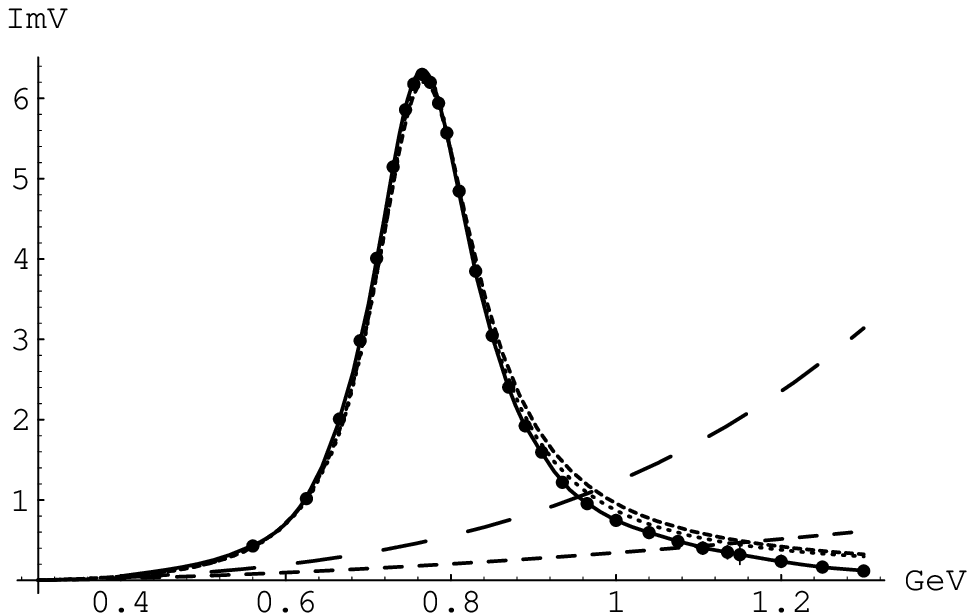}
\caption{}
\label{Fig.1}
\end{figure}

\begin{figure}
\epsfbox{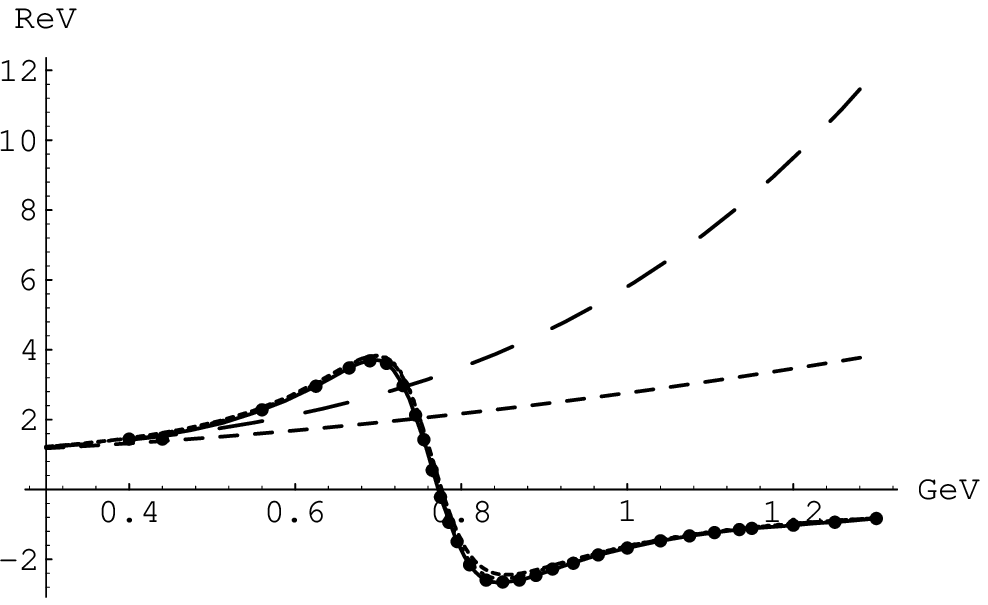}
\caption{}
\label{Fig.2}
\end{figure}
\newpage
\begin{figure}
\epsfbox{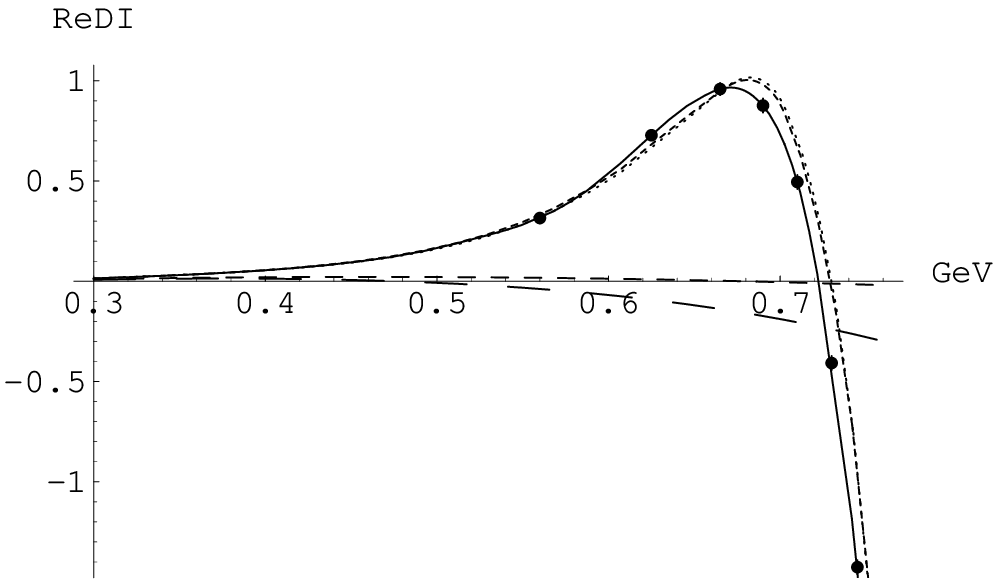}
\caption{}
\label{Fig.3}
\end{figure}

\begin{figure}
\epsfbox{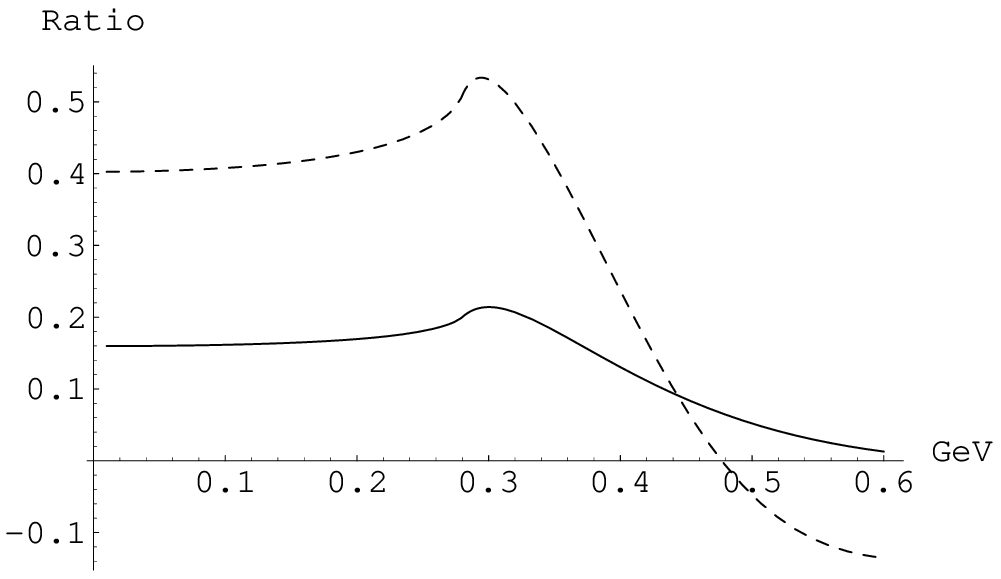}
\caption{}
\label{Fig.4}
\end{figure}

\end{document}